\documentclass[preprint]{iucr}              
     \journalcode{S}              
\usepackage{xcolor}
\usepackage{enumitem}
\usepackage{amsmath,mathtools,calc}
\usepackage{url}
\usepackage{ulem}
\usepackage[utf8]{inputenc}
\usepackage{listings}
\usepackage{multirow}

\usepackage{color,soul}

\graphicspath{{figs/}}

\definecolor{darkspringgreen}{rgb}{0.09,0.45,0.27}

\begin{document}                  



\title{TomocuPy: efficient GPU-based tomographic reconstruction with asynchronous data processing}


\cauthor{Viktor}{Nikitin}{vnikitin@anl.gov}{{address if different from \aff}}

\aff{Advanced Photon Source, Argonne National Laboratory, 60439 Lemont, IL, \country{USA}}






\maketitle                        

\begin{synopsis}
We present TomocuPy, a Python software package for fast 3D reconstruction on GPU and modern storage drives supporting parallel read-write operations. The package demonstrates significant performance gain compared to analogs and can be efficiently used when processing data during experiments requiring steering of environment conditions and acquisition schemes, or triggering data capturing processes for other measuring devices.\end{synopsis}

\begin{abstract}
Fast 3D data analysis and steering of a tomographic experiment by changing environmental conditions or acquisition parameters require fast, close to real-time, 3D reconstruction of large data volumes. Here we present a performance-optimized TomocuPy package as a GPU alternative to the commonly-used CPU-based TomoPy  package for tomographic reconstruction. TomocuPy utilizes modern hardware capabilities to organize a 3D asynchronous reconstruction involving parallel read-write operations with storage drives, CPU-GPU data transfers, and GPU computations. In the asynchronous reconstruction, all the operations are timely overlapped to almost fully hide all data management time. Since most cameras work with less than 16-bit digital output, we furthermore optimize the memory usage and processing speed by using 16-bit floating-point arithmetic.
As a result, 3D reconstruction with TomocuPy became 20-30 times faster than its multithreaded CPU equivalent. Full reconstruction (including read-write operations and methods initialization) of a $2048^3$ tomographic volume takes less than 7~s on a single Nvidia Tesla A100 and PCIe 4.0 NVMe SSD, and scales almost linearly increasing the data size. To simplify operation at synchrotron beamlines, TomocuPy provides an easy-to-use command-line interface. Efficacy of the package was demonstrated during a tomographic experiment on gas-hydrate formation in porous samples, where a steering option was implemented as a lens-changing mechanism for zooming to regions of interest.      

\end{abstract}

\section{Introduction}\label{sec:introduction}


Fast \textit{in-situ} tomographic experiments at synchrotron facilities are of great interest to various user communities including geology~\cite{butler2020mjolnir,nikitin2020dynamic}, material science~\cite{maire201620,zhai2019high}, and energy research~\cite{finegan2015operando,liu2019review}. This is because modern synchrotron light sources of the $3^{rd}$ and $4^{th}$ generation provide necessary photon flux to accommodate very fast scanning of large samples with micrometer and nanometer spatial resolution~\cite{willmott2019introduction,DeAndrade2021,nikitin2022real}. At the same time, modern detectors allow for continuous tomographic data acquisition at more than 7.7\,GB/s rate~\cite{Mokso2017,garcia2021tomoscopy} generating a series of tomographic datasets representing dynamic sample states at unprecedented high temporal resolution. 

One of the most challenging tasks nowadays is the efficient steering of such dynamic experiments. Both manual and AI-based experiment steering can be performed more efficiently utilizing full 3D reconstructed volumes rather than projection images. Reconstructed volumes are more informative and contain all information about the current sample state compared to the projection raw data or a subset of reconstructed slices. Therefore, performance of 3D tomographic reconstruction is critical when processing large amounts of data captured in a short period of time. Fast, close to real-time 3D reconstruction, will allow for AI data analysis and steering experiments, e.g., by automatic changing of environmental conditions (pressure, temperature, etc.), or by triggering data capturing processes for other measuring devices (higher resolution or ultrafast detectors). 

Besides fast reconstruction of dynamic tomography data, there is still a need for acceleration of processing large datasets, in particular, the ones obtained from detectors with large sensors or from mosaic scans. In mosaic scans, large samples are scanned at different vertical and horizontal positions to obtain a set of datasets that are then stitched together to generate one large dataset. 3D reconstructions of such large datasets can have more than 10k voxels in each dimension, yielding several Tb of data to process~\cite{vescovi2017radiography,borisova2021micrometer}. To obtain reconstruction of such datasets in a reasonable time, tomography software packages are typically adapted for high-performance computing (HPC) clusters, see for instance~\cite{hidayetouglu2020petascale} and references therein. It is common that data analysis by regular beamline users is delayed due to the lack of immediate access to such HPC clusters.

Nowadays there exist many packages for tomography data reconstruction. TomoPy \cite{gursoy2014tomopy} provides a Python interface for pre-processing tomography data and for applying filtered backprojection to recover 3D sample volumes with parallel-beam geometry. It provides implementations of different reconstruction methods, including Gridrec~\cite{Dowd:99, Rivers:12} and other~\cite{tomopyRecon}, additionally accelerated with CPU multiprocessing, Intel compiler directives, and Intel Math Kernel Library (MKL). Computational complexity for reconstructing a 3D volume is $\mathcal{O}(N^3\log N)$, assuming that the number of projection angles and volume size in each dimension are of the order of $N$.  The Gridrec implementation on computer clusters is also available~\cite{marone2017towards} and has demonstrated first steps towards on-the-fly tomography data processing. TomoPy supports Python wrappers to run reconstruction functions from other packages. One wrapper example is ASTRA Tomography Toolbox~\cite{van2015astra,van2016fast}, which is also commonly used as an independent package. The ASTRA Toolbox implements high-performance GPU primitives not only for parallel-beam tomography but also for cone-beam tomography. Besides the regular filtered backprojection method based on the summation over lines ($\mathcal{O}(N^4)$ computational complexity), the package is optimized to work with iterative reconstruction methods such as SART~\cite{andersen1984simultaneous}, SIRT~\cite{gregor2008computational}, and CGLS~\cite{scales1987tomographic}. For an iterative method, it is possible to keep all the necessary data in the GPU memory, and thereby reduce the data copy between the storage drive, CPU and GPU memory. In such cases, the performance of reconstruction is mostly limited by GPU computation speed. Another package, called UFO~\cite{vogelgesang2016real}, provides a multi-threaded, GPU-enabled and distributed data processing framework. Tomographic and laminographic reconstructions are also implemented using the regular filtered backprojection method of $\mathcal{O}(N^4)$ complexity. 

Computational complexity plays an important role when reconstructing data from large detectors, or from data obtained by stitching several projection datasets~\cite{Vescovi:18, tile}. For example, for $N=2048$ the complexity $\mathcal{O}(N^3\log N)$ becomes approximately 186 times lower than $\mathcal{O}(N^4)$. With increasing data sizes, the potential acceleration becomes higher (341 complexity lowering factor for $N=4096$, 630 - for $N=8192$, and so on), therefore it is always beneficial 
to operate with algorithms of lower computational complexity with introduction of new detectors having large sensors (e.g. 13392$\times$9528 sensor shr661 camera from SVS-VISTEK) they indeed become critical for any future tomography applications. Examples of methods with $\mathcal{O}(N^3\log N)$ complexity include the Fourier-based gridding method~\cite{beylkin1998applications} and log-polar-based method~\cite{andersson2016fast}. 
In contrast, methods implemented in ASTRA and UFO packages have $\mathcal{O}(N^4)$ computational complexity and therefore become less efficient when processing data from huge detectors.

In this work, we present a new package called TomocuPy where we combined efficient reconstruction methods and modern hardware capabilities to accelerate the whole tomographic reconstruction process including data read/write operations with storage drives, CPU-GPU data transfers, and computations on GPU. Main features of the packages include: 
\begin{itemize}
\item \textit{Optimized GPU implementation of reconstruction with low ($\mathcal{O}(N^3\log N)$) computational complexity (Fourier-based gridding method and log-polar-based method)}. The methods were developed previously, however, they have not been commonly used as a regular tool inside a tomographic package such as TomocuPy. The performance table in~\cite{andersson2016fast} reports 0.045~s log-polar-based reconstruction of 1 slice $2048\times2048$ on Nvidia GeForce GTX 770 (release date: May 30th 2013), which corresponds to 92~s for reconstructing the full volume. The reported time does not include initialization and data transfer costs. Modern GPUs are several tens of times faster than GTX 770 and reduce reconstruction times to a few seconds. 

\item \textit{Asynchronous chunk data processing where read/write operations with storage drives, CPU-GPU data transfers, and GPU computations for each chunk are timely overlapped.} It is known that one of the main bottlenecks slowing down reconstruction when using GPUs is data management. Computations on GPU may take less time than data read/write from storage drives and CPU-GPU data transfers. TomocuPy provides functionality to almost fully hide time for all data management. In this work, we optimize operation with modern storage based on Non Volatile Memory Express (NVMe) solid state disks (SSDs). They deliver unprecedented performance provided by parallelization of the read/write operations, which results in 8$\times$ acceleration compared to regular SATA SSDs~\cite{xu2015performance}. Besides computer clusters, current NVMe SSDs connected via PCIe v3 or PCIe v4 are also used in common workstations and demonstrate 3.5-7 GB/s speed for parallel operations with the disk. In this case, writing one tomographic volume of size 2048$^3$ in 32-bit precision may potentially take less than 5~s.

\item \textit{16-bit (half-precision) arithmetic.} Most detectors used for tomography have less than 16-bit digital output. It is therefore potentially possible to decrease processing data sizes and accelerate computations even more. TomocuPy implements all processing methods in both 16 and 32-bit precision. 16-bit computations decrease reconstruction sizes, accelerate computations and demonstrate acceptable accuracy for processing experimental datasets.

\item \textit{Command-line interface for reconstruction.} TomocuPy provides a command-line interface for processing tomographic datasets stored in the HDF5 format. The interface includes necessary commands and parameters for tomographic data pre-processing and reconstruction. It is also easy to extend the interface by adding new functionality with a description of parameters. 

\end{itemize}


The rest of the paper is organized as follows. In Section~\ref{sec:fast_rec} we will describe the implementation details and an easy-to-use command-line interface for processing experimental data. Section~\ref{sec:perf_acc} provides performance analysis on synthetic data and accuracy analysis on experimental data from a micro-CT synchrotron beamline and a comparison between different methods. An example of a dynamic tomography experiment at synchrotron where automatic steering was possible due to fast reconstruction provided by TomocuPy is presented in Section~\ref{sec:gh}. Conclusions and outlook are given in Section~\ref{sec:conclusions}.

\section{Fast GPU-based reconstruction with TomocuPy}\label{sec:fast_rec}
TomocuPy~\cite{tomocupy} is a Python package that provides support for fast and efficient asynchronous data management and tomographic reconstruction on Nvidia  graphics processing units (GPUs) with 16-bit or 32-bit computational precision. It implements GPU-based pre-processing steps and filtered backprojection operators, as well as optimized data transfer mechanisms among storage drives, CPU RAM memory and GPU memory. In what follows we will describe the main package features leading to fast, close to real-time, tomographic reconstruction.

\subsection{16-bit precision arithmetic}
Area detectors used for tomographic imaging incorporate an analog to digital converter (ADC) to digitize the images with 8, 10, 12 or 16-bit output. The conventional tomography reconstruction is typically performed with 32-bit floating-point operation which might be inefficient in terms of computational speed. In this work we considered 16-bit floating-point (FP16) arithmetic as an alternative to the conventional 32-bit floating-point (FP32) arithmetic. FP16 is used in many computer graphics environments to store pixels, including Nvidia CUDA, OpenGL and Direct3D. Currently, it is also gaining popularity in deep learning applications with Nvidia GPUs. Nvidia's recent Pascal architecture 
was the first GPU that offered FP16 support. FP16 arithmetic was significantly optimized for following Nvidia architectures including Volta 
and Ampere
, and became beneficial for code optimization in terms of performance and memory usage.

To adapt tomographic reconstruction in TomocuPy for FP16 computations we followed the guidance from ~\cite{ho2017exploiting} that shows different issues and opportunities with code migration to FP16. We also reviewed all mathematical operations in the code and made sure that the accuracy and correctness of computations are not lost. The accuracy can be lost when a mathematical operation is performed between large and small numbers, e.g. $1000-0.1 = 1000$ (FP16), incorrect results are obtained when multiplying two large numbers: $1000\times1000=inf$ (FP16) since the maximum representable value in FP16 precision is 65504. To address these issues, we reorganized arithmetic operations where it was possible. In places where the reorganization was not possible, we perform the operation with arguments converted to FP32 and cut the precision of the result back to FP16. As a result, we were able to decrease the total amount of memory (CPU RAM, GPU, and storage disk space) by 2 times and accelerate computations on GPU. 

\subsection{Pre-processing steps and backprojection}
Pre-processing steps in tomographic reconstruction include dark-flat field correction, taking negative logarithm of the data, one-dimensional filtering with the Shepp-Logan, Parzen or other filter. Additionally, pre-processing may include ring-removal filtering using wavelets \cite{Munch:09} or with an analytical formula by V. Titarenko~\cite{titarenko2010analytical,titarenko20161}, zinger artifacts reduction \cite{Rivers:98}, or propagation-based phase-retrieval procedure with the Paganin filter \cite{paganin2002}. To accelerate computation of all these steps we used CuPy Python library \cite{nishino2017cupy}, which is a GPU-accelerated analog of NumPy Python library. All regular linear algebra operations such as multiplication, summation, logarithm, exponent, are easily ported to CuPy library calls in 16-bit and 32-bit floating-point precision. At the time of writing this manuscript, CuPy does not support computing FFTs in 16-bit precision, moreover, 16-bit FFTs are supported in CUDA C only for the sizes that are powers of 2. Therefore, we prepared CUDA C codes for allocating 16-bit and 32-bit CUDA FFT plans at the beginning of reconstruction and executing the plans on a set of tomography slices during data reconstruction by chunks. 16-bit data are additionally padded/unpadded to the power of 2 sizes, 32-bit data are padded to the sizes represented  as $2^a\times3^b\times5^c\times7^d$ ($a,b,c,d$ are positive integer numbers) for optimal evaluation of Bluestein's algorithm \cite{Bluestein:70} in CUDA cuFFT library. 

The backprojection operator is the most computationally intensive step of the reconstruction procedure. Its direct evaluation by discretizing line integrals has computational complexity of $\mathcal{O}(N^3N_\theta)$ assuming that the sample size in each dimension and the total number of projection angles are of orders $N$ and $N_\theta$, respectively. There exist methods for fast evaluation of the backprojection operator. The most common one is based on using the Fourier-slice theorem and using Fourier transforms on unequally-spaced grids \cite{beylkin1998applications,Dowd:99}. It has complexity $\mathcal{O}(N^3\log N)$ for reconstructing 3D volumes. The same complexity demonstrates the log-polar-based method \cite{Andersson:16}. However, in comparison to the Fourier-based method where interpolation-like procedures are conducted in the frequency domain, the log-polar-based method assumes interpolation in the image domain where data are substantially less oscillatory. Therefore, the log-polar-based method demonstrates accurate reconstruction results using interpolation schemes of moderate order (linear or cubic splines), whereas the Fourier-based method has to operate with exponential, or other complex-type functions that can be approximated with only high-order polynomials. The log-polar-based method outperforms the Fourier-based method~\cite{andersson2016fast} because of the interpolation type, however, its current implementation assumes that projection data are given for equally spaced angles.
In very rare cases, e.g. during an interlaced scanning \cite{Mohan:15}, tomographic data are collected for non-equally spaced angles and the log-polar-based method is not applicable.

TomocuPy provides 3 implementations of the backprojection operator: (1) direct discretization of the backprojection line integral, (2) Fourier-based method with exponential functions for interpolation in the frequency domain, and (3) the log-polar-based method with cubic splines interpolation in the image domain. Although the direct line discretization method is not optimal, we keep it as an option since the method can be used for computing backprojection in a laminographic geometry \cite{Helfen:07} where the rotary stage is tilted with respect to the beam direction, yielding more efficient scanning of flat samples. Since the backprojection is the most computationally demanding part of reconstruction, we fully implemented it with CUDA C by writing optimized codes for FFTs, CUDA raw kernels, and data handling.
Users can easily switch between different backprojection methods depending on application.

\subsection{Asynchronous data processing}
Besides data processing on GPU, tomographic reconstruction requires data transfer operations between a storage drive, CPU RAM memory and GPU memory. Non-optimal organization of data transfers among these components, especially in GPU computing, can significantly slow down the whole reconstruction pipeline, causing GPU being idle while waiting for new data chunk transfers to complete. 
Because of non-optimal organization of data transfers, significant GPU acceleration is typically visible for iterative tomography reconstructions, where data is loaded to GPU memory and tens or hundreds of iterations are performed while keeping the whole dataset in GPU memory. One-step filtered backprojection implemented on GPU with sequential data transfers does not yield such performance gain compared to the CPU version because of relatively slow memory transfer operations.
Here we organize and optimize an asynchronous processing pipeline where all data transfers are overlapped with GPU computations. This way, time for data transfers is effectively hidden from the total computational time required by the reconstruction step. 

\begin{figure}
    \centering
    \includegraphics[width=1\textwidth]{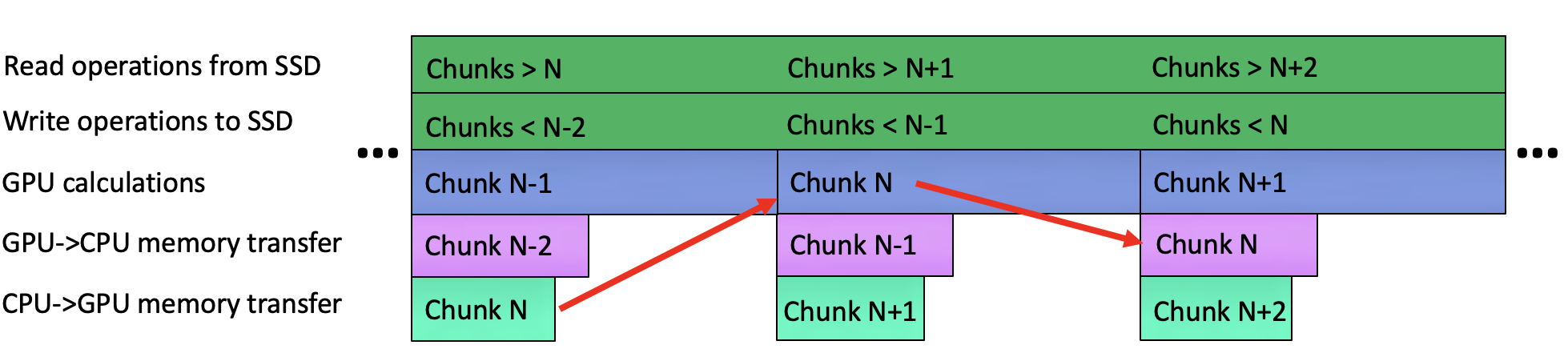}
    \caption{A scheme for asynchronous data processing by chunks where GPU reconstructions are overlapped with data transfers. }
    \label{fig:conv}
\end{figure}

Figure~\ref{fig:conv} presents a scheme of the proposed asynchronous processing pipeline for data chunks. An example of execution is as follows. When data Chunk N is loaded from a storage drive, three operations are executed simultaneously: CPU-GPU memory transfer for Chunk N, GPU computations for Chunk N-1,  GPU-CPU memory transfer for Chunk N-2. After Chunk N-2 with reconstruction is copied to CPU, a write operation is executed to dump the chunk to the storage drive. 

TomocuPy implements this optimal asynchronous pipeline in two levels. First, independent Python threads are started for (1) reading data chunks from the storage drive into a Python data queue object and for (2) writing reconstructed chunks from another Python queue object to the storage drive disk. Both queue objects are stored in CPU RAM memory. The size and the number of threads for each queue are defined based on the system characteristics. To maximize performance of parallel read/write operations we work with Intel\textsuperscript{\textregistered} SSD D7-P5510 Series PCIe 4.0 NVMe drives. These drives work on high-end parallel data paths for faster operations than regular SAS/SATA HDDs or SSDs, protocols of which are based on CPU cycles and are not designed to handle severe data loads.

Second, independently on read/write operations with storage drives, we overlap CPU-GPU data transfers with GPU computations by using CUDA Streams
. CuPy interface allows organizing the concurrent execution of streams directly within the Python code, without writing CUDA C code. It also allows for direct allocation of pinned GPU memory, which is necessary to run data transfers and GPU computations concurrently. To implement the overlap, the pinned memory on CPU and device memory on GPU should be both allocated for 2 data chunks and 2 reconstruction chunks. Three CUDA streams are running simultaneously by switching between chunks, the first stream performs a data copy to the first chunk of the pinned memory, followed by the transfer to the first chunk of GPU memory. The second stream performs GPU computations on the second data chunk in GPU memory (whenever it is available)  and places the result to the second reconstruction chunk in GPU memory. The third stream executes a data transfer from the first reconstruction chunk in GPU memory to the first pinned memory chunk for reconstruction. The chunk is then copied to the queue for further writing to the storage drive. After processing each chunk, all streams synchronize and switch the chunk id (0 or 1) they operate with.

\begin{figure}
    \centering
    \begin{tabular}{c}
         \small{Asynchronous execution (40 ms timeline)}\\
         \includegraphics[width=0.95\textwidth]{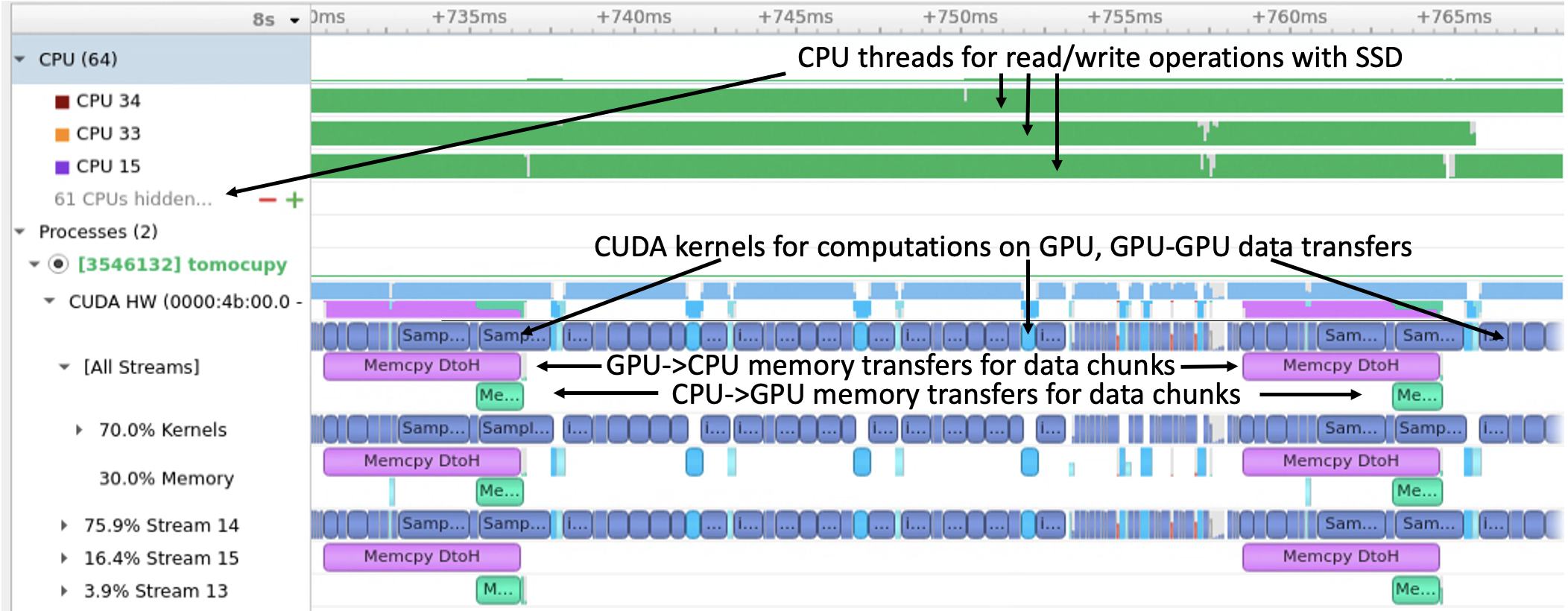}\\[0.3cm]
         \small{Sequential execution (160 ms timeline)}\\
        \includegraphics[width=0.95\textwidth]{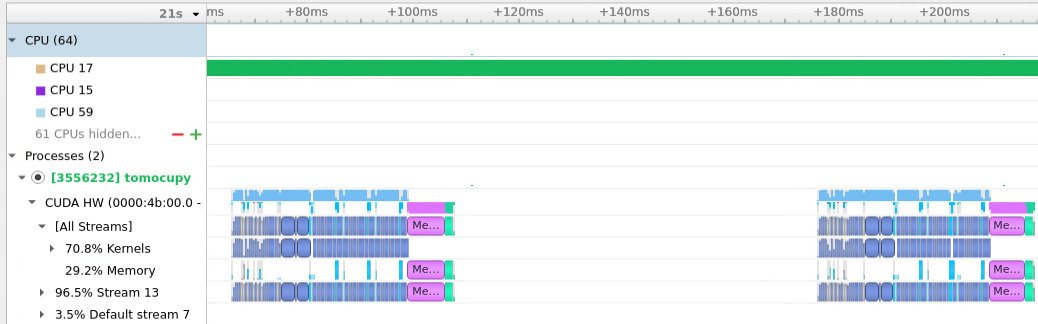}
    \end{tabular}
    \caption{Timeline view report from Nvidia Nsight System tool for asynchronous and sequential execution of reconstruction with TomocuPy. Reconstruction was performed for a $2048\times2048\times2048$ dataset with the log-polar-based method and FP32 arithmetic.}
\label{fig:Nsight}
\end{figure}

Figure \ref{fig:Nsight} shows the timeline view report from the Nvidia Nsight System performance analysis tool 
that  demonstrates a comparison between the asynchronous and sequential execution types. The test was performed for reconstructing a $2048\times2048\times2048$ dataset with the log-polar-based method and FP32 arithmetic. The timeline view for the asynchronous execution is shown for 40 ms. During this time, we observe continuous data read/write operations with NVMe SSD, so as continuous CUDA kernel execution and GPU-GPU memory transfers. We also observe 2 GPU-CPU and 2 GPU-CPU memory transfers for data chunks. Since all operations are overlapped in time by using Python CPU threads and CUDA Streams, the total reconstruction time in this case can be approximately estimated only by GPU computations. In turn, the timeline view for the sequential execution with 1 CPU thread and 1 CUDA Stream shows 160 ms running time without any overlap. During this period, about 60\% of the time is spent for read/write operations with an NVMe SSD in 1 thread, the remaining 40\% is used for GPU computations and memory transfers between CPU and GPU. The left panels of both reports also show that the total GPU utilization consists of about 70\% CUDA kernel execution and 30\% memory transfers. Total time for reconstructing a $2048\times2048\times2048$ dataset with the asynchronous execution, as measured by the Nvidia Nsight System performance analysis tool, is approximately 2.5 times lower than for the sequential execution (8 s vs 21 s).

\subsection{Command-line interface}
To simplify the execution of tomographic reconstructions with TomocuPy, we have developed a command-line interface wrapping Python classes with processing functions. The command can be executed in a Unix terminal and accepts a list of parameters to customize the reconstruction procedure. The executable file is installed as a part of the whole TomocuPy package by using pip or conda install.  
An example of a command line for running in an Anaconda environment and reconstructing one full tomographic dataset stored as an HDF5 file is as follows:  
\begin{verbatim}
$ tomocupy recon --file-name /data/testing_131.h5 --rotation-axis 1224 
--reconstruction-type full
\end{verbatim}
where the reconstruction parameters are submitted with the syntax \verb|--<parameter> <value>|. The list of all available parameters can be obtained by running \verb|tomocupy recon -h|. The description of some parameters (approx. 20\% of the full list) looks as follows:

\begin{verbatim}
  --binning {0,1,2,3}   Reconstruction binning factor (default: 0)
  --dtype {float32,float16}
                        Data type used for reconstruction. Note float16 works 
                        with power of 2 sizes. (default: float32)
  --fbp-filter {shepp,parzen}
                        Filter for FBP reconstruction (default: parzen)
  --max-read-threads MAX_READ_THREADS
                        Max number of threads for reading by chunks (default: 1)
  --max-write-threads MAX_WRITE_THREADS
                        Max number of threads for writing by chunks (default: 4)
  --nsino NSINO         Location of the sinogram used to find rotation axis 
                        (0 top, 1 bottom) (default: 0.5)
  --nsino-per-chunk NSINO_PER_CHUNK
                        Number of sinograms per chunk. Use larger numbers with GPUs 
                        with larger memory. (default: 8)
  --rotation-axis ROTATION_AXIS
                        Location of rotation axis (default: -1.0)
  --rotation-axis-auto {manual,auto}
                        How to get rotation axis: auto calculate ('auto'), 
                        or manually ('manual') (default: manual)
  --reconstruction-algorithm {fourierrec,lprec,linesummation}
                        Reconstruction algorithm (default: fourierrec)
  --reconstruction-type {full,try}
                        Reconstruct full data set. (default: try)
  --start-row START_ROW
                        Start slice (default: 0)
  --end-row END_ROW
                        End slice (default: -1)

\end{verbatim}

A general reconstruction procedure typically consists of two steps: 1) reconstruction of 1 slice for different rotation centers and saving reconstructed tiff files with names corresponding to these centers (parameter \verb|--reconstruction-type try|). Users then open all files and select the rotation center by scrolling through different tiff files. The selected rotation center is then entered (\verb|--rotation-axis 1224|) and used for the full reconstruction (\verb|--reconstruction-type full|). TomocuPy also provides an automatic center search option (\verb|--rotation-axis-auto auto|) by using SIFT algorithm~\cite{lowe1999object} to find shifts between 0 and (flipped) 180 degrees projections.

Command-line interface for TomocuPy was developed to assure compatibility with the command-line interface TomoPy-cli (\url{https://tomopycli.readthedocs.io}) for CPU-based reconstruction. TomoPy-cli uses TomoPy package \cite{gursoy2014tomopy} as a backend and implements an efficient workflow for processing tomographic data files (tiff, HDF5) from storage drives. Both packages, TomocuPy and TomoPy-cli, has the same syntax for passing parameters. They also provide the same names for most of parameters, except method-specific parameters, such as \verb|--dtype, --max-read-threads|, etc. Likewise, the input/output format, file names are identical. It is therefore not complicated to switch between two packages and compare performance and quality of reconstruction results.

It is important to note that a multi-GPU version of tomographic reconstruction is straightforward to implement because in the parallel beam geometry reconstruction is done independently for different slices through the volume. TomocuPy provides parameters \verb|--start-row| and \verb|--end-row| for specifying the range of slices for reconstruction, therefore multi-GPU reconstruction can be performed, for instance, by setting environment variable \verb!CUDA_VISIBLE_DEVICES! associated with the GPU number and running daemon processes in bash for each subset of slices,
\begin{verbatim}
CUDA_VISIBLE_DEVICES=0 tomocupy recon --file-name test2048.h5 \
                       --reconstruction-type full --start-row 0 --end-row 512 &
CUDA_VISIBLE_DEVICES=1 tomocupy recon --file-name test2048.h5 \ 
                       --reconstruction-type full --start-row 512 --end-row 1024 &
CUDA_VISIBLE_DEVICES=2 tomocupy recon --file-name test2048.h5 \ 
                       --reconstruction-type full --start-row 1024 --end-row 1536 &
CUDA_VISIBLE_DEVICES=3 tomocupy recon --file-name test2048.h5 \ 
                       --reconstruction-type full --start-row 1536 --end-row 2048 &
wait
\end{verbatim}

Since the processes are independent the total performance will be limited only by the storage and system bus speed for data transfers. 

\section{Performance and accuracy analysis}\label{sec:perf_acc}

To check reconstruction quality that TomocuPy package demonstrates when processing experimental datasets, we collected tomographic projections for a sample consisting of 20-40~$\mu$m glass beads packed in a kapton tube with 4 mm diameter. The measurements were performed at the bending magnet micro-CT beamline 2-BM~\cite{nikitin2022real} of the Advanced Photon Source. The beamline was adjusted for using the pink beam (polychromatic X-ray beam reflected from a grazing mirror) cutting energies higher than 30 keV, and with additional 6 mm glass filtering of low energies. Projections were acquired by a CMOS detector Oryx 5.0 MP Mono 10GigE,  2448$\times$2048 chip size, 3.45~$\mu$m pixel size, made by Teledyne FLIR LLC. The detector used a 2x magnification infinity corrected objective by a Mitutoyo resulting in 1.725~$\mu$m pixel size. The lens was focused to a 50~$\mu$m Crytur LuAG:Ce scintillator converting X-rays to visible light.

Tomographic projections were acquired in fly scanning mode, while the sample was continuously rotated for a 180 degrees interval. In total 2048 projections of the size 2048$\times$2048 (cropped field of view for the detector) were collected with 0.05~s exposure time per projection, yielding 1.7~min total acquisition time. 
The reconstruction procedure was performed by using 3 reconstruction algorithms implemented in TomocuPy, and by using Gridrec method from TomoPy (with TomoPy-cli interface for data pre-processing and transfers):
\begin{itemize}
    \item FourierRec - Fourier-based method with exponential-function interpolation in the frequency domain~\cite{beylkin1998applications}, computational complexity $\mathcal{O}(N^3\log N)$
    \item LpRec - Log-polar-based method with cubic interpolation in the space domain \cite{andersson2016fast}, computational complexity $\mathcal{O}(N^3\log N)$
    \item LineRec - Direct discretization of the line integral with linear interpolation for computing backprojection, computational complexity $\mathcal{O}(N^4)$
    \item Gridrec - TomoPy implementation of the Fourier-based method~\cite{Dowd:99}, computational complexity $\mathcal{O}(N^3\log N)$ 
\end{itemize}
where the computational complexity is calculated assuming the number of projection angles, and the object size in each dimension are of the order of $N$. All methods employed the commonly-used Parzen filter for implementing filtered backprojection.

\begin{figure}
    \centering
         \includegraphics[width=\textwidth]{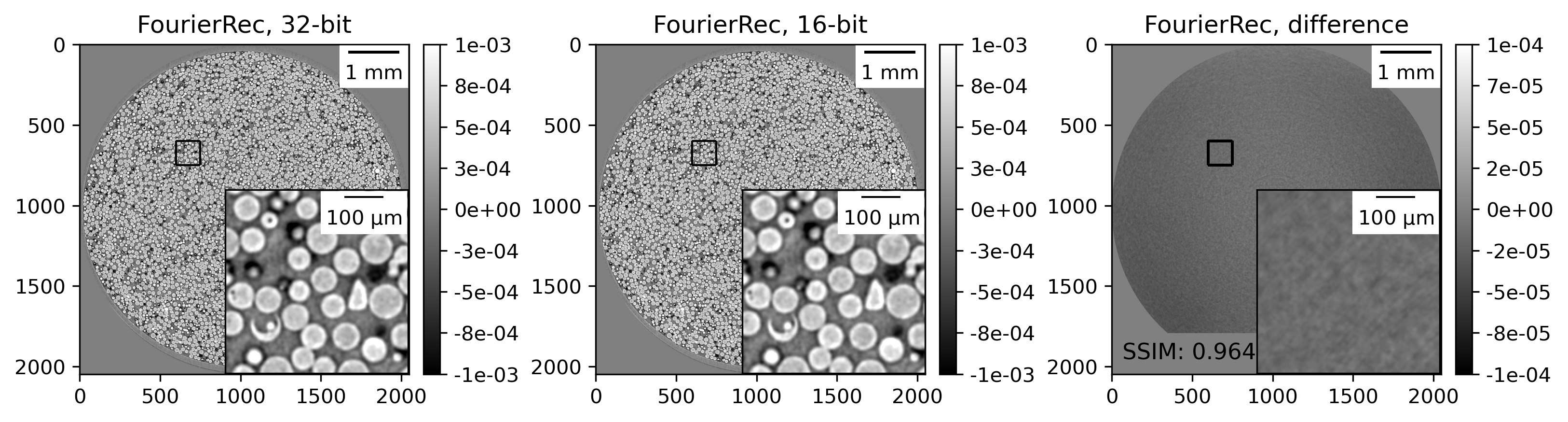}\\
         \includegraphics[width=\textwidth]{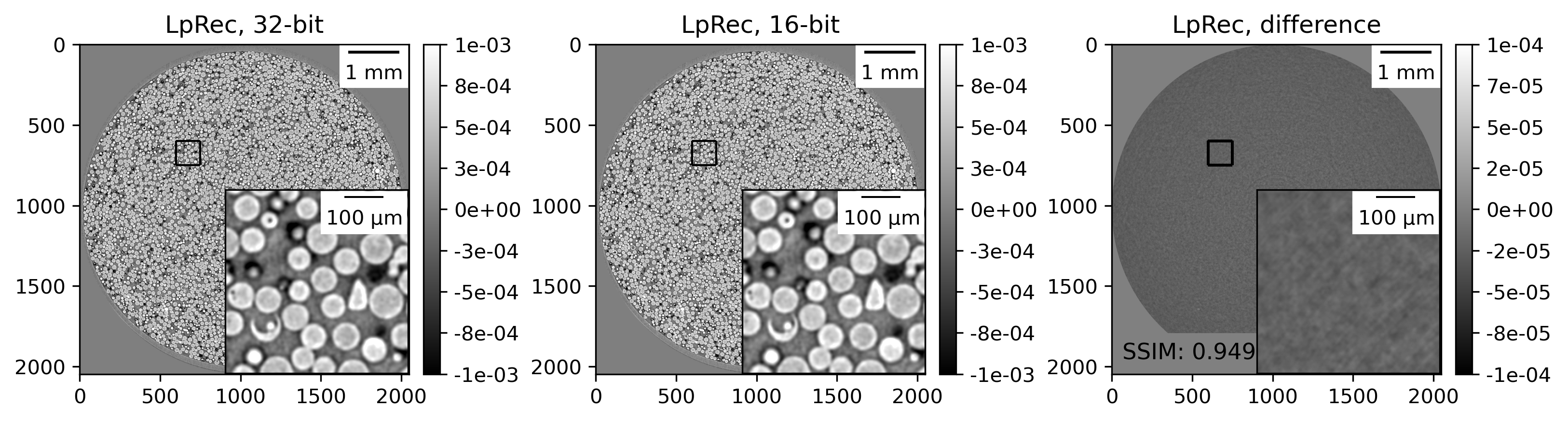}\\
         \includegraphics[width=\textwidth]{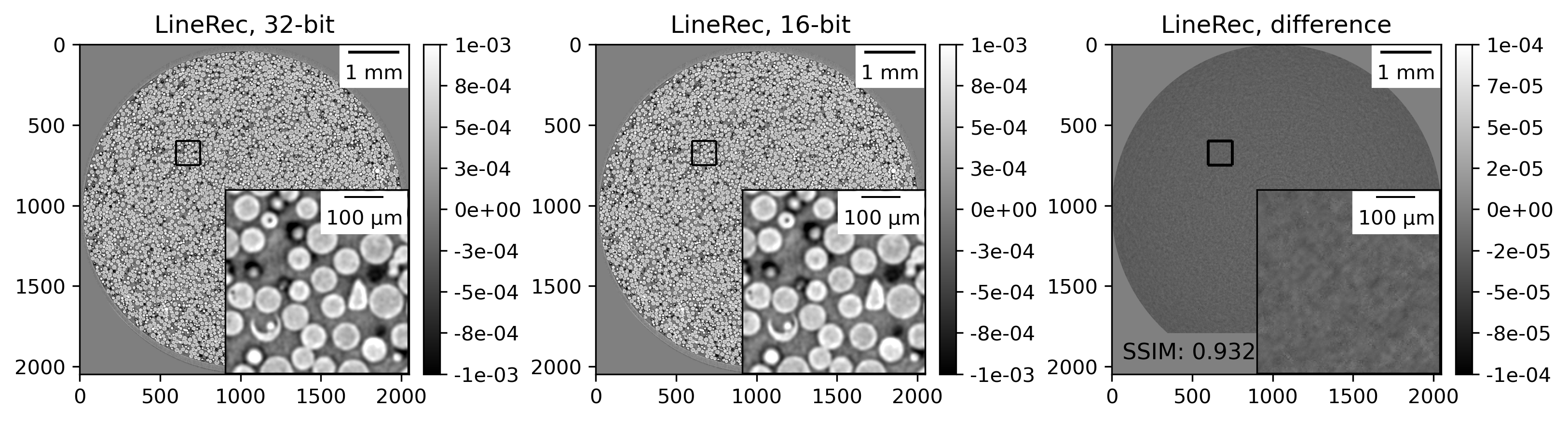}\\
    \caption{Comparison of micro glass beads reconstructions for 16 and 32-bit floating-point precision arithmetic. Inset plots show 10x zooming to the region marked by the black square. Colorbar range for the difference plot is 10 times smaller than for reconstructions.}
\label{fig:diff_fp32_16}
\end{figure}

Figure~\ref{fig:diff_fp32_16} presents a comparison for reconstructions using TomocuPy (FourierRec, LpRec, LineRec) with 32 and 16-bit floating-point arithmetic. Each image shows one reconstructed slice using different methods, together with insets showing 10x zoom-in to the region marked with the black rectangle. Visually, reconstructions for FP32 and FP16 look the same. The right part of the figure shows the difference between them, so as calculated Structural Similarity Index (SSIM)~\cite{wang2004image} quantifying image quality degradation. SSIM is higher than 0.93 for all methods, which confirms high quality of FP16 results. Note that the data were collected in 12-bit precision, i.e. in maximum precision for most of tomographic detectors used in fast imaging. For additional confirmation, we checked accuracy with 16-bit synthetic Shepp-Logan phantom datasets generated as described in Section~7 from \cite{andersson2016fast}.  The results confirmed that the error of FP16 computations is negligible compared to FP32. We therefore can conclude that 16-bit arithmetic is enough for processing tomographic data, and all reconstructed volumes can be stored using twice lower amount of memory.       

As a second quality test, we compared TomocuPy reconstructions with the ones produced by the Gridrec method implemented in TomoPy. Gridrec is a Fourier-based method, i.e. is using the Fourier-slice theorem and fast evaluation of Fourier transforms on unequally spaced grids. The difference with its TomocuPy equivalent, called FourierRec, is in the interpolation kernels used for data re-gridding in the frequency domain, and oversampling factors for frequencies. TomoPy implementation of Gridrec does not include oversampling, therefore regular reconstruction contains phase wrapping artifacts. In order to minimize these artifacts, additional padding of sinograms is typically performed before the filtered backprojection operation~\cite{marone2012regridding}. FourierRec includes oversampling by a factor of 2 and accuracy controls in computing the backprojection integral. A detailed accuracy analysis for evaluating backprojection with the Fourier and Log-polar-based methods for the Shepp-Logan phantom sample can be found in~\cite{andersson2016fast}. In the paper, filtered versions of the Shepp-Logan phantom, as well as corresponding projection data, are computed analytically and therefore directly used for evaluating the backprojection error for different methods. Based on the fact that the FourierRec method is the method with the highest accuracy in computing backprojection (based on the accuracy tests from ~\cite{andersson2016fast}), we will present results for other methods in comparison to FourierRec. Additionally, all reconstruction methods implemented in TomocuPy involve data padding for the filtering operation, which allows for suppressing artifacts when processing samples not fitting into the detector field of view. 

\begin{figure}
    \centering
         \includegraphics[width=0.64\textwidth]{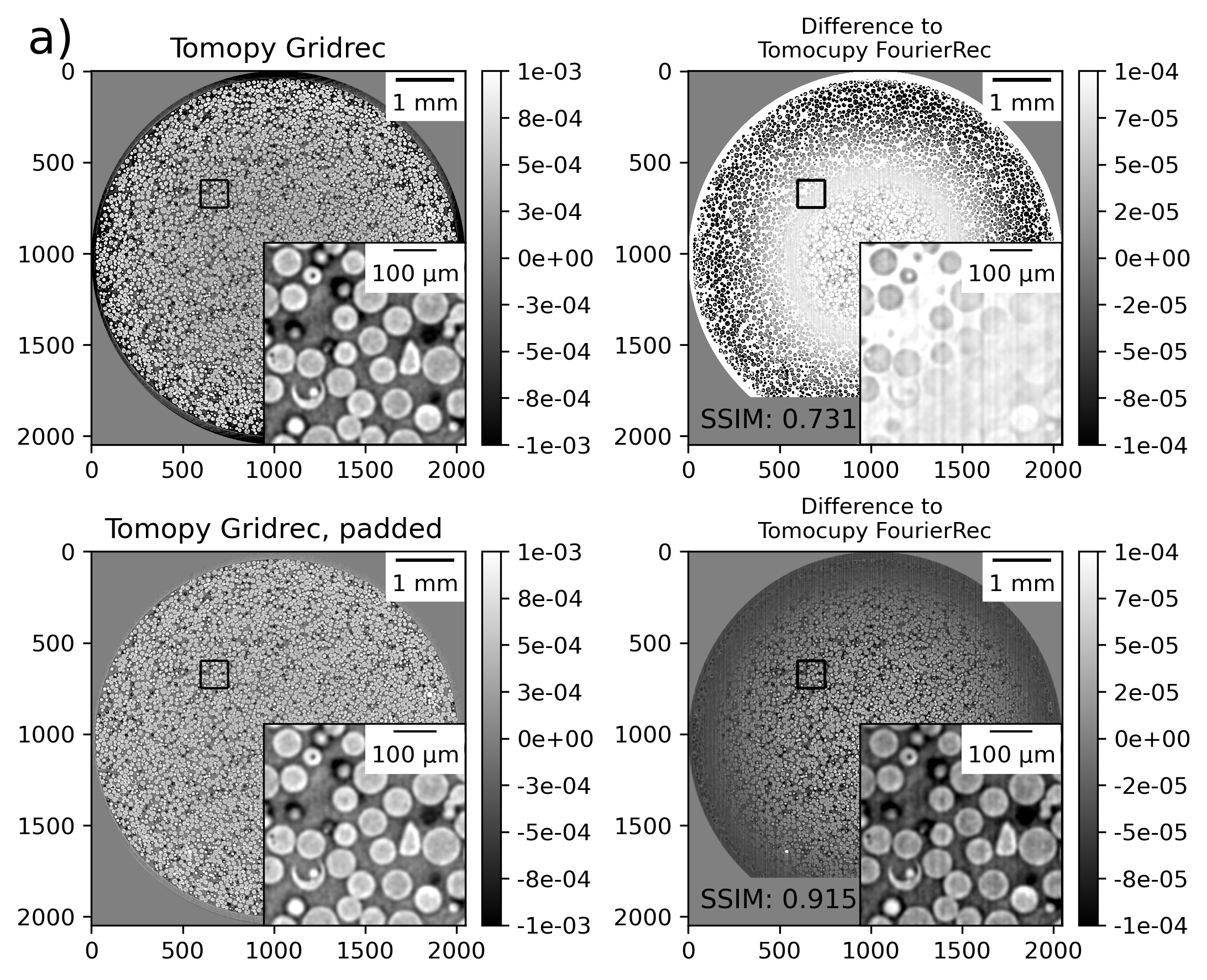}\hspace{0.4cm}
         \includegraphics[width=0.32\textwidth,trim=264 0 0 0,clip]{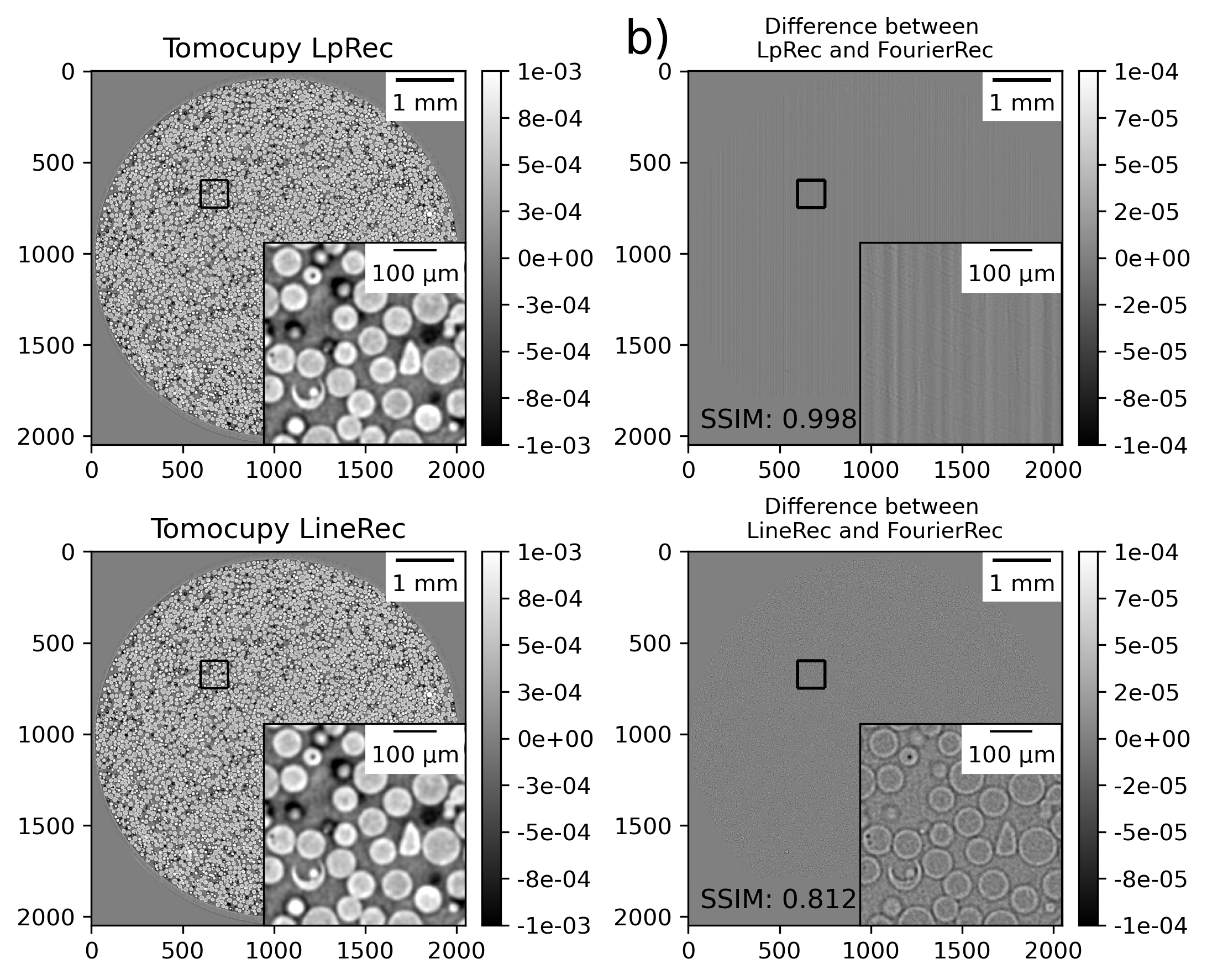}\\
    \caption{Comparison reconstruction results in 32-bit floating-point precision: a) for TomocuPy FourierRec and TomoPy Gridrec (with and without padding of sinograms) methods, and b) between TomocuPy FourierRec, LpRec and LineRec. Inset plots show 10x zooming to the region marked by the black square. Colorbar range for the difference plots is 10 times smaller than for reconstruction. }
\label{fig:diff_to_tomopy}
\end{figure}

In Figure~\ref{fig:diff_to_tomopy}a we show the difference between reconstructions between Gridrec from TomoPy and FourierRec from TomocuPy. One can see that regular TomoPy Gridrec reconstruction (top row) has errors in low-frequency components, visible as amplitude changes in the regions close to the borders. SSIM is relatively low (0.731). In turn, reconstruction with additional sinogram padding (bottom row) does not have visually observed amplitude changes, however, the difference to TomocuPy FourierRec still highlights errors in low frequencies. Despite the errors in low frequencies, TomoPy is still commonly used for reconstructing tomographic data because information given by high frequencies (small features) is more important in several applications, and it is accurately recovered with SSIM=0.915. Note again that the accuracy of the methods implemented in TomocuPy was checked using analytical expressions for the Shepp-Logan phantom and its projection data~\cite{andersson2016fast}.

In Figure~\ref{fig:diff_to_tomopy}b we provide the difference images between LpRec and FourierRec (top), and between LineRec and FourierRec (bottom). One can observe very high accuracy of the LpRec method where cubic interpolations to and from log-polar coordinates are done in the image domain. The LineRec method is implemented with linear interpolation in the image domain, thus errors in high-frequency components are clearly visible. Structural similarity index for these two methods are 0.998 and 0.812, respectively.

For the performance analysis, TomocuPy package was tested using synthetic HDF5 datasets of different sizes. Synthetic datasets were generated for $N$ 16-bit tomographic projections with  $N\times N$ detector sizes, where $N$ ranges from 512 to 16384. Reconstructed volumes $(N\times N\times N)$ were obtained as sets of tiff files in 16-bit and 32-bit precision. Note that the selected projection data sizes do not satisfy the Crowther sampling criterion stating that the number of angles should be $\frac{\pi}{2}N\approx\frac{3}{2} N$~\cite{crowther1970reconstruction}. In tomographic experiments this criterion is typically relaxed, and reconstruction results with acceptable quality are demonstrated for a significantly lower number of angles, e.g. equal to $N$ or $3/4N$.

For completeness, we also analyzed the performance of TomoPy-cli package, where all pre-processing steps and Gridrec reconstruction were accelerated using multi-threaded CPU functions and Intel\textsuperscript{\textregistered} Math Kernel Library. Recall that both TomoPy-cli and TomocuPy command-line interfaces have almost the same set of parameters and in most cases can be easily interchanged. 

Performance tests were performed on a machine with Intel Xeon\textsuperscript{\textregistered} Gold 6326 CPU @ 2.90GHz, 1 TB DDR4 3200 memory, 1 Nvidia Tesla A100 with 40 GB memory, and Intel SSD D7-P5510 Series PCIe 4.0 NVMe disks of total capacity 84 TB. Installed software included Python 3.9, CuPy 10.4.0, Nvidia CUDA toolkit 11.6, and Intel Math Kernel Library Version 2022.1 (only for fast CPU-based computations in TomoPy).  

Table~\ref{tab:performance} shows the dataset dimensions used to test the performance of tomocuPy methods (FourierRec, LpRec, LineRec) and the CPU-based TomoPy Gridrec. 

Table~\ref{tab:performance1} shows the total time to reconstruct the test datasets listed in Table~\ref{tab:performance} using TomocuPy's  FourierRec and LpRec methods with FP16 and FP32 precision, TomocuPy's LineRec with FP32 precision, as well results for TomoPy-cli package where all pre-processing steps and Gridrec method for reconstruction are executed in FP32 precision.

\begin{table}[]
    \centering
    \setlength{\tabcolsep}{2pt}
    \begin{tabular}{|c|c|c|c|c|c|c|c|c|}
         \hline
         Test dataset& 1 & 2 & 3 & 4 & 5 & 6\\
         \hline
         Size in each & \multirow{2}{*}{512} & \multirow{2}{*}{1024} &\multirow{2}{*}{2048} &\multirow{2}{*}{4096} &\multirow{2}{*}{8192} &\multirow{2}{*}{16384}\\
         dimension, $N$&&&&&&\\
         \hline
         Raw data size on & \multirow{2}{*}{128MB} & \multirow{2}{*}{1GB} & \multirow{2}{*}{8GB} & \multirow{2}{*}{64GB} & \multirow{2}{*}{512GB} & \multirow{2}{*}{4TB}\\
         SSD, (8-bit)&&&&&& \\
         \hline
         Reconstruction size & \multirow{2}{*}{256(512)MB} & \multirow{2}{*}{2(4)GB} & \multirow{2}{*}{16(32)GB} & \multirow{2}{*}{128(256)GB} & \multirow{2}{*}{1(2)TB} & \multirow{2}{*}{8(16)TB}\\
         on SSD, 16(32)-bit&&&&&& \\
         \hline
    \end{tabular}
    
    \caption{Dataset dimensions used to test the performance of tomocuPy methods (FourierRec, LpRec, LineRec) and TomoPy Gridrec. }
    \label{tab:performance}
\end{table}    

\begin{table}[]
    \centering
    \setlength{\tabcolsep}{2pt}
    \renewcommand{\arraystretch}{1.2}
    \begin{tabular}{|c|c|c|c|c|c|c|c|c|}
         \hline
         Test dataset& 1 & 2 & 3 & 4 & 5 & 6\\
         \hline
         FourierRec, 16-bit & {$4.2\!\times\!10^{-1}$s} & {$1.7\!\times\!10^{0}$s} & {$1.1\!\times\!10^{1}$s} & {$8.2\!\times\!10^{1}$s} & {$6.8\!\times\!10^{2}$s} & {$6.7\!\times\!10^{3}$s}\\
         \hline
         FourierRec, 32-bit & {$6.3\!\times\!10^{-1}$s} & {$2.4\!\times\!10^{0}$s} & {$1.5\!\times\!10^{1}$s} & {$1.2\!\times\!10^{2}$s} & {$9.8\!\times\!10^{2}$s} & {-}\\
         \hline
         LpRec, 16-bit &{$3.1\!\times\!10^{-1}$s} & {$1.1\!\times\!10^{0}$s} & {$6.4\!\times\!10^{0}$s} & {$5.2\!\times\!10^{1}$s} & {$4.9\!\times\!10^{2}$s} & {$5.2\!\times\!10^{3}$s}\\
         \hline
         LpRec, 32-bit & {$4.3\!\times\!10^{-1}$s} & {$1.4\!\times\!10^{0}$s} & {$7.3\!\times\!10^{0}$s} & {$5.9\!\times\!10^{1}$s} & {$5.5\!\times\!10^{2}$s} & {-}\\
         \hline
         LineRec & {$6.5\!\times\!10^{-1}$s} & {$5.5\!\times\!10^{0}$s} & {$8.0\!\times\!10^{1}$s} & {$1.3\!\times\!10^{3}$s} & {$2.1\!\times\!10^{4}$s} & {-}\\
         \hline
         TomoPy Gridrec, 32-bit & {$3.7\!\times\!10^{0}$s} & {$1.5\!\times\!10^{1}$s} & {$1.4\!\times\!10^{2}$s} & {$1.7\!\times\!10^{3}$s} & {$1.6\!\times\!10^{4}$s} & {$1.7\!\times\!10^5$s$^*$} \\
         \hline
    \end{tabular}

    $^*$ estimated by using a lower number of chunks.
    
    \caption{Total time to reconstruct the datasets listed in Table~\ref{tab:performance} using TomocuPy's  FourierRec and LpRec methods with FP16 and FP32 precision, TomocuPy's LineRec with FP32 precision, and TomoPy-cli package where all pre-processing steps and Gridrec method for reconstruction are executed in FP32 precision. Raw data (16-bit) and reconstructed volumes (16 or 32-bit) both have sizes $N \times N \times N$, assuming the number of projection angles equals to $N$. Reconstruction time includes all parts of the processing pipeline (reading HDF5 data chunks from NVMe SSD, writing reconstructed tiff files to NVMe SSD, CPU-GPU transfers, and all CPU/GPU computations). }
    \label{tab:performance1}
\end{table}

There are several observations from Table~\ref{tab:performance1}. First, all methods allow working with very large data sizes (up to several TB on SSD), which is useful for processing data from the detectors with large sensors or from mosaic tomographic scans. 

Second, we observe that FP16 computations not only reduce data sizes but also accelerate the reconstruction step by approximately 10\% and 30\% for LpRec and FourierRec, respectively. Double memory size reduction allowed processing data for $N=16384$, which was not possible for 32-bit precision due to the GPU memory limit.
We think that the difference between acceleration factors for the two methods (10\% and 30\%) is caused by the implementation of cubic B-spline interpolation procedures in the log-polar-based method, see~\cite{andersson2016fast} for details. Hard-wired linear interpolation is implemented in GPU texture memory and works with lower than 32-bit precision due to the texture reading access organization. As a result, the performance of read/write texture access with interpolation might not be significantly different for 16 and 32-bit precision. 

Third, the table shows that for large data sizes the fastest TomocuPy method (LpRec) outperforms CPU-based TomoPy implementation of Gridrec by a factor of 33 and 29 for 16 and 32-bit precision, respectively. It has been already demonstrated that GPU is more efficient than CPU for tomographic data reconstruction, see Table 1 in \cite{andersson2016fast}. However, additional time for CPU-GPU data transfers, so as read/write operations with storage drives, concealed this efficiency. With the asynchronous execution proposed in this work, the benefits of using GPU became clearly visible.

Finally, it is important to note that computational complexity of algorithms is crucial in accelerating reconstruction algorithms. Computational complexity of all algorithms presented in  Table~\ref{tab:performance1} is $O(N^3\log N)$, except for LineRec that has complexity $O(N^4)$. Although LineRec is also optimized and works via the asynchronous execution, its reconstruction time for large data sizes is higher than the one for the CPU-based TomoPy reconstruction. We expect to see similar performance behaviour when working with other GPU-based implementations, such as ASTRA toolbox wrapper inside TomoPy~\cite{pelt2016integration} or UFO package~\cite{vogelgesang2016real}, where the backprojection method has computational complexity of $\mathcal{O}(N^4)$. With such complexity, total reconstruction time for large data volumes will be mostly estimated by the GPU processing time, since time for all data transfers increases linearly with data sizes. More accuracy and performance comparisons between the Fourier-based, log-polar, ASTRA Toolbox, and other methods can be found in~\cite{andersson2016fast}.

In the previous section, we mentioned that multi-GPU reconstruction can be performed by setting environment variable \verb!CUDA_VISIBLE_DEVICES! associated with the GPU number and running daemon processes in bash for subsets of slices. For demonstration, we executed reconstruction on 1 node of Polaris \footnote{https://www.alcf.anl.gov/polaris} supercomputer of Argonne Leadership Computing Facility. Compared to the workstation used for preparing Table 1, a Polaris node is equipped with a more powerful processor (AMD EPYC Milan series) and 4 Tesla A100 GPUs with the SXM connection interface (not PCI Express) and having high-speed HBM memory architecture. The storage is also based on NVMe PCIe v4 SSDs. From Table~\ref{tab:performance2} one can see that time scaling with increasing the number of GPUs is almost linear for the FourierRec method. In turn, LpRec method demonstrates an overhead when the number of GPUs is increased from 2 to 4 (e.g., $1.9\!\times\!10^{2}$s vs {$1.3\!\times\!10^{2}$s} for test dataset 5). We explain this overhead by the fact that GPU computations for LpRec are faster than for FourierRec and thus time for data management becomes more significant. Indeed, in this case 4 processes associated with GPUs compete with each other for the storage and system bus used for data transfers.

\begin{table}[]
    \centering
    \setlength{\tabcolsep}{2pt}
    \renewcommand{\arraystretch}{1.2}
    \begin{tabular}{|c|c|c|c|c|c|c|c|}
         \hline
         Test dataset &\multicolumn{3}{c|}{5}&\multicolumn{3}{c|}{6}\\
         \hline
         Number of GPUs &1 & 2 & 4 & 1 & 2 & 4\\
         \hline
         FourierRec, 16-bit &{$6.2\!\times\!10^{2}$s} &{$3.1\!\times\!10^{2}$s}&{$1.6\!\times\!10^{2}$s} &{$5.7\!\times\!10^{3}$s} & {$3.0\!\times\!10^{3}$s}  & {$1.6\!\times\!10^{3}$s} \\
         \hline
         LpRec, 16-bit &{$3.6\!\times\!10^{2}$s} & {$1.9\!\times\!10^{2}$s} & {$1.3\!\times\!10^{2}$s}&{$3.7\!\times\!10^{3}$s} & {$1.9\!\times\!10^{3}$s} & {$1.1\!\times\!10^{3}$s} \\
         \hline
    \end{tabular}
    
    \caption{Total time to reconstruct the test datasets 5 ($N=8192$) and 6 ($N=16384$) by using 1 node of the Polaris supercomputer (4 Tesla A100 GPUs with HBM memory, storage consisting of NVMe PCIe v4 SSDs). Reconstruction parameters are the same as in Table \ref{tab:performance1}. }
    \label{tab:performance2}
\end{table}

To provide additional performance comparison that could be relevant for the readers, in Table~\ref{tab:performance3} we report reconstruction times on a regular workstation equipped with 1 NVidia Quadro RTX 4000 GPU and an NVMe SSD connected via PCI Express v3.0. This workstation is less expensive and therefore affordable for most tomographic beamline users. The table shows that such workstation still demonstrates favorable performance results when processing tomographic data.     

\begin{table}[]
    \centering
    \setlength{\tabcolsep}{2pt}
    \renewcommand{\arraystretch}{1.2}
    \begin{tabular}{|c|c|c|c|c|c|c|c|c|}
         \hline
         Test dataset& 1 & 2 & 3 & 4\\
         \hline
         FourierRec, 16-bit & {$2.1\!\times\!10^{0}$s} & {$4.6\!\times\!10^{0}$s} & {$2.2\!\times\!10^{1}$s} & {$1.8\!\times\!10^{2}$s} \\
         \hline
         LpRec, 16-bit &{$2.3\!\times\!10^{0}$s} & {$4.6\!\times\!10^{0}$s} & {$1.9\!\times\!10^{1}$s} & {$1.5\!\times\!10^{2}$s}\\
         \hline
    \end{tabular}
    
    \caption{Total time to reconstruct test datasets 1-4 ($N=512\dots4096$) by using a regular workstation(1 Quadro RTX 4000, 1 NVMe PCIe v3 SSD). Reconstruction parameters are the same as in Table \ref{tab:performance1}.}
    \label{tab:performance3}
\end{table}

\section{Dynamic tomography experiment with steering}\label{sec:gh}
To briefly demonstrate efficacy of the TomocuPy package for processing data from a dynamic experiment with steering we considered \textit{in-situ} multi-resolution study of gas-hydrate formation inside porous media. The setup of the experiment can be found in~\cite{nikitin2020dynamic,nikitin2021dynamic}, multi-resolution scanning of gas-hydrates with an automatic lens-changing mechanism of the Optique Peter system\cite{OP} is described in~\cite{nikitin2022real}. 

 
 
The whole sample was represented as a cylinder with height 2~cm and diameter 0.5~cm. For low-resolution scanning of the middle part of the sample, 1200 projections were acquired with 0.04~s exposure time per projection, which together with dark/flat fields acquisition yielded 50~s per one scan.  High-resolution scans with 5x lens were acquired with 1800 angles per scan and 0.15~s exposure time. The reconstruction procedure involved dark-flat field correction, ring removal, taking negative logarithm, and filtered backprojection by the log-polar-based method. Additional phase-retrieval filtering~\cite{paganin2002} was applied for processing high-resolution data to enhance gas hydrate contrast in local tomography imaging.  

For the steering demonstration, we were monitoring the gas-hydrate formation process in low spatial resolution (1.1x lens), detected regions with fast water flows happening spontaneously, and automatically zoomed-in to these regions for higher resolution  (5x lens) scanning. Such automatic steering allowed us to capture the initiation and evolution of the gas-hydrate growth process inside the pore initially filled with water. 

\begin{figure}\label{fig:gh}
    \centering
        \includegraphics[width=0.7\textwidth]{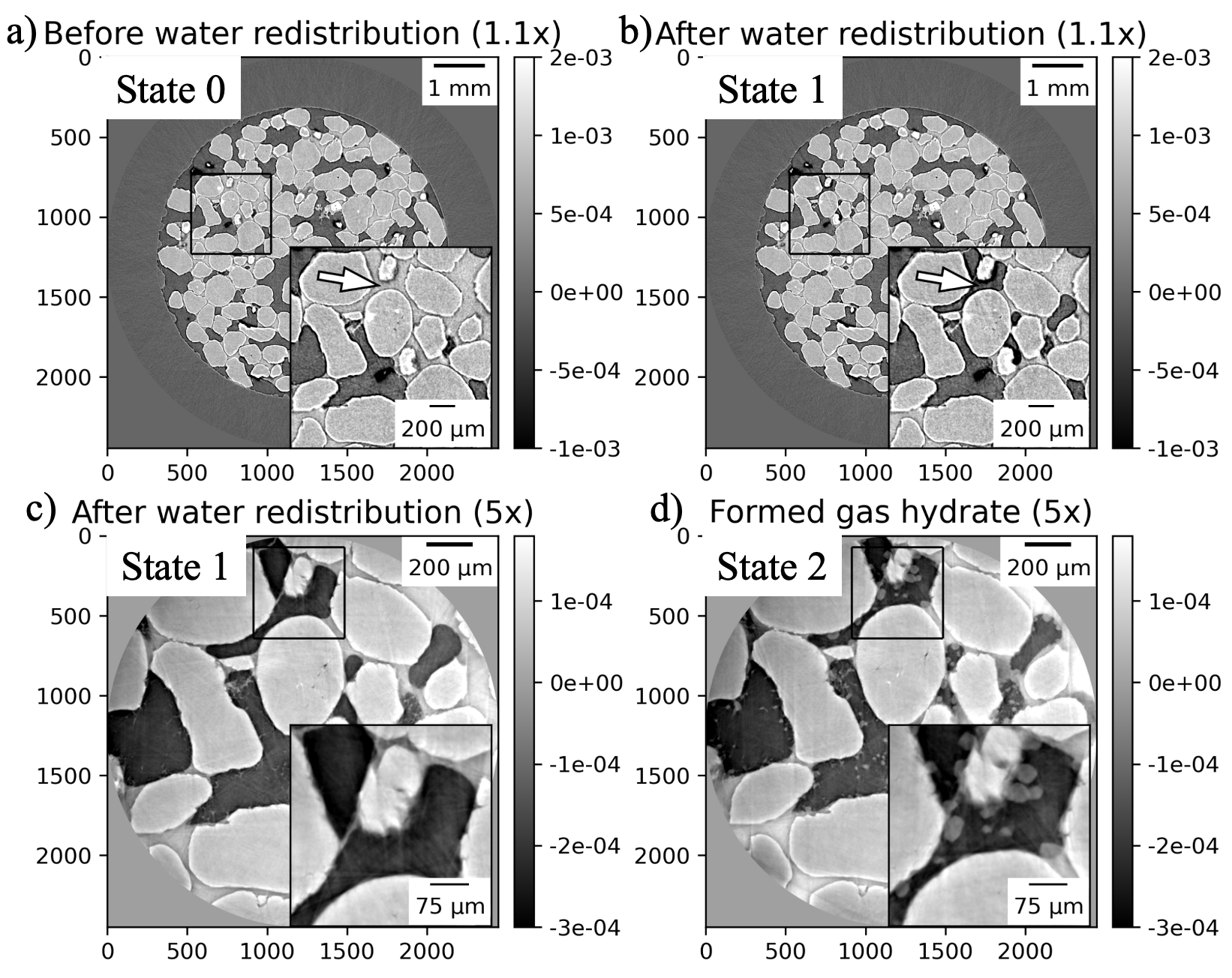}
    \caption{Gas-hydrate formation experiment with automatic steering (zooming to a region of interest with water outflow): a,b) sample states in low spatial resolution before (State 0) and after (State 1) water redistribution, respectively; c) the region of interest in high spatial resolution after water redistribution (State 1), d) the region of interest with formed gas hydrate (State 2).  Bright color corresponds to sand grains and water solution, dark gray/black - to methane gas, light gray in high resolution images - to gas hydrate.}
\label{fig:setup}
\end{figure}

Detecting regions with fast water flows was done after reconstructing full data volumes with TomocuPy and comparing them with the ones from the previous sample state by taking element-wise difference. Reconstruction and the region of interest detection took approximately 12 s, which is much less than the total scan time (50~s). Therefore the steering engine had sufficient time to select the appropriate region for high-resolution scanning with 5x lens. Figure~\ref{fig:gh}a,b shows slices through reconstructed volumes in low resolution for the sample state before and after water redistribution. In this figure, bright color corresponds to sand grains and water solution, and dark gray/black - to methane gas. The region with water outflow is marked by white arrows. Immediately after the low-resolution scan, this region was detected and scanned with high resolution (Figure~\ref{fig:gh}c), where the hydrate structure formed on the water-gas interface can be observed in light gray color. The region was further continuously scanned until the final state (the end of experiment time) shown in Figure~\ref{fig:gh}d. 

In Table~\ref{tab:steering} we provide a part of the timeline for the gas-hydrate formation experiment with steering. Note, that some actions, such as scanning for one state and reconstruction for another, are overlapped in time. Although this study is far from real-time, it still demonstrates an example of automatic steering implementation. This dynamic study can be potentially accelerated. First, one can switch to fast data acquisition (e.g., with pink beam). Second, data transfers to the processing machine can be avoided by broadcast data directly to the CPU RAM or GPU memory. Third, TomocuPy reconstructions can be done with binning (Table~\ref{tab:performance1} shows that reconstruction of $1024^3$ takes about 1 second). Finally, the motorized lens changing mechanism of the Optique Peter microscope system can be replaced by a pneumatic mechanism, which will spend less than a second to switch the lens as opposed to 5 seconds for the current motorized system.

\begin{table}[]
    \centering
    {
    \begin{tabular}{|l|l|}
        \hline
         Time & Action \\
         \hline
         00:00 - 00:50 & Low-resolution scan for State 0\\
         \hline
         00:50 - 00:53 & State 0 data transfer to the processing machine \\
         \hline 
         00:50 - 01:40 & Low-resolution scan for State 1\\
         \hline
         00:53 - 01:01 & Reconstruction for State 0 \\
         \hline
         01:40 - 01:43 & State 1 data transfer to the processing machine\\
         \hline 
         01:40 - 01:55 & Next low-resolution scan (not finished \\&because the ROI found earlier) \\
         \hline
         01:43 - 01:51 & Reconstruction for State 1  \\
         \hline
         01:51 - 01:55 & Automatic selection of the ROI by taking the \\
         & difference between State 0 and State 1 reconstructions \\& (both are in CPU memory)  \\
         \hline
         01:56 - 02:01 & Automatic lens change with the Optique Peter \\& system and moving the sample stack motor\\& to have the ROI in the middle of the field of view  \\
         \hline
         02:02 - 06:34 & High-resolution scan for State 1\\
         \hline
         06:34 - 06:38 & State 1 data transfer to the processing machine\\
         \hline
         06:38 - 10:10 & High-resolution scan for State 2\\
         \hline
         06:34 - 06:45 & Reconstruction for State 1, including phase retrieval\\
         \hline
         ...& ...\\
         \hline
    \end{tabular}
    }
    
    \caption{Timeline for the gas-hydrate experiment with a steering demonstration.}
    \label{tab:steering}
\end{table}
\section{Conclusions and Outlook}\label{sec:conclusions}
By developing TomocuPy package we have shown that full tomography reconstruction from a standard detector (2k$\times$2k sensor size), including all read/write operations with storage drives and initialization functions, can be done in less than 7~s on a single Nvidia Tesla A100 and NVMe PCIe v4 SSD. The asynchronous data processing almost completely hides the CPU-GPU data transfers time and read/write operations with storage drives are optimized for parallel operations. Additionally, switching to 16-bit floating point arithmetic decreased memory usage and processing times without significant reduction in reconstruction quality. The package is publicly available at \url{https://readthedocs.org/projects/tomocupy}.

Performance tests showed almost linear time scaling with increasing data sizes up to 16384$^2$ slices. The linear scaling is due to efficient TomocuPy algorithms with low computational complexity ($N^3\log N$), which becomes beneficial when working with modern detectors having large sensors. Full processing time to reconstruct a $16384^3$ volume on 1 GPU is approximately 1.5 h, and can be decreased with adding GPUs because tomography slices are processed independently. For comparison, a CPU-based reconstruction with an Intel\textsuperscript{\textregistered} Xeon Gold takes approximately 47 h, i.e. requires at least 33 computing nodes and a fast GPFS storage to demonstrate TomocuPy performance. 
Reconstruction on one node of Polaris supercomputer with 4 more powerful GPUs and fast NVMe storage took about 20 mins. 
We note that Tesla A100 (40 GB) has enough memory to process 16384$^3$. If GPU memory is not enough, the TomocuPy reconstruction engine automatically switches to using Unified Memory~\cite{chien2019performance} and process data by automatically transferring data parts to and from CPU RAM memory. However, since automatic CPU-GPU data transfers with Unified Memory typically show low performance, we still plan to optimize reconstruction algorithms by also processing each slice by chunks asynchronously. Specifically, we will optimize 2D FFTs and interpolation functions in FourierRec (interpolation to a polar grid in the frequency domain) and LpRec (interpolation to polar and log-polar grids in the space domain). 2D FFT can be represented as a combination of 1D FFTs and thus computed by chunks on GPUs. Evaluation of interpolations to irregular grids can be also done by splitting all grid points into chunks that are independently processed by GPU. We expect that by using an optimized asynchronous pipeline implemented with CUDA Streams, the overhead for CPU-GPU data transfers will be negligible, which will allow us to process huge datasets on GPU in a reasonable time. Similar pipelines can be constructed for chunked read/write operations with storage drives if data do not fit into CPU RAM memory.

Fast 3D tomographic reconstruction with TomocuPy opens new possibilities for automatic steering \textit{in-situ} experiments.  As a first application, we considered a geological experiment for gas hydrate formation in porous media, where the initiation of the formation process after water redistribution was captured in high resolution inside a large sample. As the next step, we plan to conduct gas-hydrate experiments with varying cooling temperature based on the sample state. It has been shown with acoustic measurements that temperature cycles affect the hydrate growth speed~\cite{dugarov2019laboratory} and new tomography measurements may provide more details about this process. The steering mechanism could have a wide range of applications not only in geosciences but also in materials science, environmental science, and medical research. We plan to study the crack formation process inside different materials. The crack will be initiated using a load cell while low-resolution projection data are continuously captured and reconstructed. The deformed regions of interest will be measured with high resolution to monitor the crack initiation in details. Additionally, we plan to vary pressure according to the sample state obtained from reconstructions.

Although in this work we have demonstrated steering with sub-minute temporal resolution, most of the listed applications require imaging with sub-second resolution and corresponding sub-second reconstruction speed. As Table~\ref{tab:performance1} shows, such reconstruction speed with TomocuPy can already be achieved for the data volumes that are smaller than $1024^3$. To steer most dynamic experiments there is no need for data reconstruction in high resolution, which means that TomocuPy can potentially be used with real-time dynamic experiments where detector data are slightly cropped or binned. However, the package needs a couple of adjustments for that. First, we need to organize streaming data processing as in \cite{nikitin2022real} where data is transferred directly from the detector to the processing computer over the high-speed network and where  data capturing to the storage drive is done on-demand. Second, 3D reconstructed volumes will be generated by TomocuPy in real-time and thus have to be immediately analyzed (segmented, classified, etc.). We envision that fast Machine Learning-based techniques (probably running on an independent GPU) should optimize data analysis and generate quick automatic feedback to the acquisition system. For instance, in \cite{tekawade2022real} the authors have recently shown an example of real-time tomographic data analysis that can be adapted for different applications. 

TomocuPy package can be extended by adding new processing and reconstruction methods. New methods implemented with Python NumPy library are directly adapted for GPU computations by switching to Python CuPy library.   Currently TomocuPy provides GPU implementations only of the one-step filtered backprojection, which is explained by the aim of having reconstructions as fast as possible. Iterative reconstruction schemes are significantly slower but they still can be added to process data more efficiently. Iterative schemes, especially the ones where slices are not reconstructed independently (e.g, 3D Total Variation regularization), can be implemented more efficiently with asynchronous GPU reconstruction and CPU-GPU data transfers. The same holds also for laminography reconstruction where data chunks can be organized not only in data slices but also in projections. If data volumes are too large then the asynchronous pipeline should also include read/write operations with the storage drive.

TomocuPy package is in routine use at the micro-CT 2-BM and nano-CT 32-ID beamlines. Because of its easy-to-use command-line interface that has almost the same commands as the one used by TomoPy, the package has quickly become friendly for beamline users. Data reconstruction for most experiments is currently done during the experiment beamtime.

\ack{Acknowledgements}

The author thanks Francesco De Carlo and  Anton Duchkov for their insightful comments and help with improving the paper structure, Alex Deriy and Pavel Shevchenko for help in conducting the gas-hydrate formation experiment.  
This research used resources of the Advanced Photon Source, a U.S. Department of Energy (DOE) Office of Science user facility at Argonne National Laboratory and is based on research supported by the U.S. DOE Office of Science-Basic Energy Sciences, under Contract No. DE-AC02-06CH11357.


\referencelist     

\end{document}